\RequirePackage{fix-cm}
\documentclass[smallcondensed]{svjour3}     
\smartqed  
\usepackage{natbib}
\usepackage{graphicx}
\usepackage{mathrsfs} 
\usepackage{bm,amsmath}
\usepackage{amsfonts} 
\usepackage{algorithmicx}
\usepackage{algorithm}  
\usepackage{hyperref}
\usepackage{enumitem}
\usepackage{algpseudocode}  
\usepackage{threeparttable}
\usepackage{booktabs}
\usepackage{multirow}
\usepackage{pgfplots}
\usepackage{subfig}
\usepackage{xspace}
\usepackage{bm}
\usepackage{float}

\newcommand\subparagraph{%
	\@startsection{subparagraph}{5}
	{\parindent}
	{3.25ex \@plus 1ex \@minus .2ex}
	{-1em}
	{\normalfont\normalsize\bfseries}}
	
\usepackage{titlesec}
\usepackage{tablefootnote}
\newcommand{\Mat}[1]{\bm{#1}}
\usepackage{authblk}
\titleformat{\paragraph}[runin]
{\normalfont\bfseries\itshape}{\thesubsection}{1ex}{}
\titlespacing*{\section}{0pt}{10dd plus 1pt minus 0pt}{5dd}
\titlespacing*{\subsection}{0pt}{10dd plus 1pt minus 0pt}{5dd}
\titlespacing*{\paragraph}{0pt}{4pt plus 2pt minus 1pt}{5dd}
\makeatletter
\DeclareRobustCommand\onedot{\futurelet\@let@token\@onedot}
\def\@onedot{\ifx\@let@token.\else.\null\fi\xspace}

\def\eg{\emph{e.g}\onedot} 
\def\ie{\emph{i.e}\onedot}

\def\wrt{w.r.t\onedot} 

\makeatother

\begin{document}
	
	\title{Scalable Attack on Graph Data by Injecting Vicious Nodes}

	\titlerunning{Scalable Attack on Graph Data by Injecting Vicious Nodes}        
	\author{\textbf{Jihong Wang} \and \textbf{Minnan Luo} \and \textbf{Fnu Suya} \and \textbf{Jundong Li} \and \textbf{Zijiang Yang} \and \textbf{Qinghua Zheng}
	}

	\authorrunning{Jihong Wang et al.} 
	
	\institute{Jihong Wang \and Minnan Luo \and Qinghua Zheng \at
              Xi’an Jiaotong University, China \\
              \email{wang1946456505@stu.xjtu.edu.cn, minnluo@xjtu.edu.cn, tjlu@mail.xjtu.edu.cn} 
           \and 
           Fnu Suya \and Jundong Li  \qquad \qquad \qquad \qquad \qquad  Zijiang Yang \at
             University of Virginia, USA \qquad \qquad \qquad \qquad  Western Michigan University, USA \\
             \email{\{suya, jundong\}@virginia.edu} \qquad \quad \ \  \email{zijiang.yang@wmich.edu}
}

	
	\date{Received: date / Accepted: date}

	\maketitle
	
	\begin{abstract}
	Recent studies have shown that graph convolution networks (GCNs) are vulnerable to carefully designed attacks, which aim to cause misclassification of a specific node on the graph with unnoticeable perturbations. However, a vast majority of existing works cannot handle large-scale graphs because of their high time complexity. Additionally, existing works mainly focus on manipulating existing nodes on the graph, while in practice, attackers usually do not have the privilege to modify information of existing nodes. In this paper, we develop a more scalable framework named Approximate Fast Gradient Sign Method (AFGSM) which considers a more practical attack scenario where adversaries can only inject new vicious nodes to the graph while having no control over the original graph. Methodologically, we provide an approximation strategy to linearize the model we attack and then derive an approximate closed-from solution with a lower time cost. To have a fair comparison with existing attack methods that manipulate the original graph, we adapt them to the new attack scenario by injecting vicious nodes. Empirical experimental results show that our proposed attack method can significantly reduce the classification accuracy of GCNs and is much faster than existing methods without jeopardizing the attack performance. 
	
	\end{abstract}
	
	\section{Introduction}
	Graphs are widely used to model various types of real-world data and many canonical learning tasks such as classification, clustering, and anomaly detection have been widely investigated for the graph-structured data \citep{bhagat2011node,tian2014learning,perozzi2014focused,tang2016node}. In this paper, we focus on the task of node classification. Recently, to solve the node classification problem, graph convolution networks (GCNs) have gained a surge of research interests in the data mining and machine learning community because of their superior prediction performance~\citep{pham2017column,cai2018comprehensive,monti2017geometric}. However, recent research efforts showed that various graph mining algorithms (e.g., GCNs) are vulnerable to carefully crafted adversarial examples, which are ``unnoticeable'' to humans but can cause the learning models to misclassify some target nodes~\citep{zugner2018adversarial,dai2018adversarial,bojchevski2018adversarial}. The vulnerabilities of these learning algorithms can lead to severe consequences in security-sensitive applications. For example, GCNs are often used in the risk management area to evaluate the credit level of users~\citep{dai2018adversarial,akoglu2015graph,bolton2001unsupervised}, as user-user information is often used in this context, it provides ample opportunities for criminals to increase their credit score by connecting to high-credit users. In this paper, we focus on assessing the robustness of graph convolution networks against adversarial attacks. Different from existing efforts that directly manipulate the original graph, we investigate a more realistic attack scenario of injecting vicious nodes and develop a scalable solution to tackle the problem. 
	
	\paragraph{{\bfseries Limitations of Current Approaches.}} There are two common issues of existing works on attacking GCNs ~\citep{zugner2018adversarial, dai2018adversarial, bojcheski2018adversarial}: the scalability issue and the explicit assumption that adversaries can easily manipulate existing nodes on the graph. First, GCNs are usually applied in large-scale graphs in various domains~\citep{ying2018graph,hamilton2017inductive,chen2018fastgcn}, which puts a high demand on the scalability of the underlying attack models. However, existing efforts~\citep{zugner2018adversarial,dai2018adversarial,bojcheski2018adversarial} cannot be easily generalized to handle large-scale graphs. Second, in actual life, attackers may not be able to manipulate existing nodes in a graph. For example, GCNs are often used to classify users on social websites like Twitter or Weibo for content recommendation by exploring the friendship graph of users. But attackers usually have no ability to manipulate existing users on these websites. To achieve the purpose of attack, a simple way is to register some new accounts on these websites and enable these accounts to establish connections with existing users, \eg, following other users or making comments on the same posts. Despite that, existing attack models seldomly consider this new attack scenario. The aforementioned two limitations motivate us to investigate the following research problem: \emph{how to effectively and efficiently manipulate the prediction results of GCNs on a specific node by injecting vicious nodes to a large graph?}    
		\begin{figure}[t] 
		\hspace{-0.5cm}
		\centering
		
		\includegraphics[width=0.9\columnwidth]{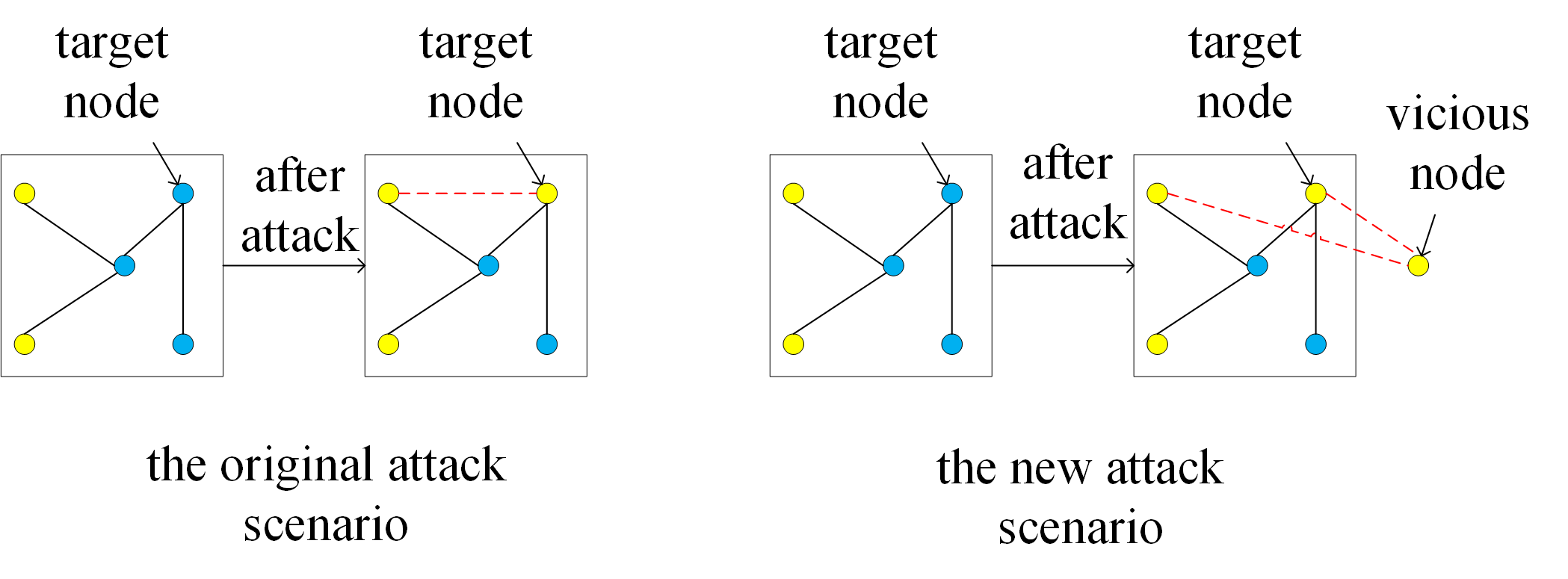}
		
		\caption{Comparison between the attack scenario in existing literature and the new attack scenario considered in this paper.}
		\label{attacks}
	\end{figure}
	\paragraph{{\bfseries Challenges.}} There are three challenges in the new attack scenario, where the first two are general challenges for devising efficient attacks on GCNs while the third one is a unique challenge of the new attack scenario we consider.

	\begin{itemize}[noitemsep,topsep=0pt]
	    \item  {\bfseries Discreteness.} Different from images which can be approximately regarded as a continuous field as the value of pixels can be any integers between 0 and 255, graph-structured data is often depicted in the discrete domains. Thus existing attack \citep{dong2018boosting,szegedy2013intriguing}strategies that are widely used in other domains (\eg, computer vision) cannot be directly applied. The proposed solution needs to well handle the potential combinatorial problem efficiently. 
	    \item  {\bfseries Poisoning Attack.} 
	    Unlike the clear separation between training and test data in other domains, the node classification task is often conducted in a transductive setting, where the test data (without ground-truth labels) is also considered in the training phase. Because of this, when the test data is manipulated, the graph is also dynamically updated. Therefore, it is important to propose attack strategies that remain effective after the model is retrained on the manipulated data. This leads to a bi-level optimization problem which is often computationally expensive to solve.
	    \item {\bfseries High complexity of Existing methods.} We can adapt existing methods to the new attack scenario. However, they all fail to scale to large graphs as their time complexities are very high (the complexity analysis is shown in Section \ref{sec:complexity}). Considering that the node classification task is usually conducted on large graphs in practice, it is important to develop an efficient method that can be scaled to real-world large graphs.
    \end{itemize} 
\paragraph{{\bfseries Contributions.}} With the above-mentioned challenges, we propose a novel Approximate Fast Gradient Sign Method (AFGSM), which can modify and inject vicious nodes efficiently in the new attack setting. Specifically, our contributions can be summarized as follows:
\begin{itemize}[noitemsep,topsep=0pt]

	\item {\bfseries A New Attack Scenario.} We consider a more practical attack scenario where adversaries can only inject vicious nodes to the graph while the original nodes on the graph remain unchanged. 
	\item {\bfseries Adapting Existing Attacks to New Scenario.} We adapt and carefully tune the exiting attacks to our new attack scenario and adopt these attacks as the baselines for comparison.
    \item {\bfseries A More Efficient and Effective Algorithm.} We propose a new attack strategy named Approximate Fast Gradient Sign Method (AFGSM), which can generate adversarial perturbations much more efficiently than the baseline attacks while maintaining similar attack performance. 
    \item {{\bfseries Extensive Evaluation.}} We empirically illustrate the effectiveness of our method on five benchmark datasets and also test on two state-of-the-art graph neural networks and one unsupervised network embedding model. 
	
\end{itemize}

\section{Related Work}
Adversarial examples are extensively studied in the image classification task and recently researchers also show its existence in graph-related problems. Therefore, we will discuss related works in both the image domain and the graph domain.    

\paragraph{{\bfseries Attack on Images.}}
\citet{szegedy2013intriguing} first demonstrate the vulnerability of deep learning models to adversarial examples using L-BFGS method and attributes the existence of adversarial examples to high non-linearity of deep models. Later on,~\citet{goodfellow2014explaining} propose an efficient Fast Gradient Sign Method (FGSM) and instead demonstrates that adversarial examples exist because deep learning models are linear in nature. Carlini et al.~\citep{carlini2017towards} propose a stronger C\&W attack that breaks heuristic defenses that are effective against adversarial examples generated using L-BFGS method. Madry et al.~\citep{madry2017towards} propose the Projected Gradient Attack (PGD) and successfully break defenses that are effective against FGSM attacks. C\&W and PGD attacks are commonly adopted as benchmarks for evaluating the robustness of new defenses as non-certified defenses can be easily evaded by considering some variants of the two attacks~\citep{athalye2018obfuscated}. However, these attacks cannot be applied to our setting because of the discrete nature of graph data. 

\paragraph{{\bfseries Attack on Graphs.}} Some earlier attacks on graphs focus on modifying the graph structure.~\citet{Chaoji2012Recommendations} add edges to maximize the content spreading in social network platforms such as Twitter. Researchers also reveal that the shortest path of a graph can be changed by slightly perturbing its structure~\citep{israeli2002shortest,Phillips1993The}.~\citet{Cs2014PageRank} aim to maximize the PageRank score of a target node in a network with structure manipulation. However, all these attacks are not designed for graph learning algorithms (\eg, graph neural networks or node embedding algorithms such as DeepWalk).

Recently, researchers also demonstrate the vulnerability of graph learning algorithms.~\citet{Chen2017Practical} inject noise to a bipartite graph that represents DNS queries to mislead the result of graph clustering. However, their attack is generated through manual effort based on attacker's domain knowledge.~\citet{zhao2018data} study the poisoning attacks on multi-task relationship learning, but based on an assumption that the sampled nodes are i.i.d. within each task, which does not hold for the node classification task.~\citet{dai2018adversarial} study evasion attacks on node classification and graph classification problems. However, their perturbation is only limited to the edges of the graph, while attackers can benefit more by additionally manipulating features of nodes (shown in Section~\ref{sec:strict limits}).~\citet{zugner2018adversarial} propose Nettack, which manipulates both edges and features of the graph with a greedy approach. In addition, the authors propose an efficient method to calculate the constraint condition on the perturbations to ensure the generated perturbations are ``unnoticeable''.~\citet{bojcheski2018adversarial} study poisoning attack on unsupervised node embeddings by borrowing ideas from matrix perturbation theory to maximize the loss of DeepWalk \citep{perozzi2014deepwalk} and change the embedding outcome.~\citet{sun2019node} study the new attack scenario by injecting vicious nodes to perturb the graph using a reinforcement learning strategy. However, the complexity of reinforcement learning ~\citep{watkins1992q} is pretty high thus cannot scale to large graphs. ~\citet{zugner2019adversarial} propose a data poisoning attack named Meta-attack based on meta-learning, which can reduce the overall classification accuracy of the GCN by only perturbing small fraction of the training data.

Note that, all the aforementioned attacks except \citep{sun2019node} on graph learning algorithms focus on changing edges or features of existing nodes and are not designed for injecting vicious nodes to the graph. As shown in Section \ref{sec:evaluation}, directly adapting these attacks to the new attack scenario is not promising due to high complexity and we are motivated to devise a new attack strategy tailored for the vicious node setting with a low computational cost.

\section{Problem Definition}
\subsection{Notation and Preliminary}
In this section, we first introduce the notations used throughout this paper and preliminaries of GCNs. Here, following the standard notations in the literature \citep{kipf2016semi,zhou2018graph,wu2019comprehensive}, we assume that $\mathcal{G}=\left(\mathcal{V},\mathcal{E}, \mathcal{F}\right)$ denotes an undirected attributed network (graph) with $n$ nodes (\eg, $v_i\in\mathcal{V}$), $m$ edges (\eg, $e_{ij}=(v_i, v_j)\in\mathcal{E}$), and $d$ attributes (features) (\eg, $f_i\in\mathcal{F}$).
The features of these $n$ nodes are given by $\Mat{X}=\left[\Vec{x}_1,\Vec{x}_2,\cdots,\Vec{x}_n\right]^\top\in\{0,1\}^{n\times d}$, where $\Vec{x}_i\in\{0,1\}^{d}$ denotes the feature for the $i$-th node $v_i$. 
The adjacency matrix $\Mat{A}\in\{0,1\}^{n\times n}$ contains the information of node connections, where each component $\Mat{A}_{ij}$ denotes whether the edge $e_{ij}$ exists in the graph. 
Here, we represent the graph as $\mathcal{G}=\left(\Mat{A},\Mat{X}\right)$ for simplicity.
Note that only a limited number of nodes possess label information in many real-world scenarios and we denote these nodes as $\mathcal{V}_{L}$, where each node $v_i\in\mathcal{V}_{L}$ is affiliated with the label $c_v\in\mathcal{C}$.

Following the well-established work on node classification \citep{kipf2016semi}, the probability of classification with GCN is formulated as:
\begin{equation}
\label{eq2}
\Mat{Z} = f_{\bm{\theta}}(\mathcal{G}) = softmax\left(\hat{\Mat{A}}\sigma\left(\hat{\Mat{A}}\Mat{X}\Mat{W}^{(1)}\right)\Mat{W}^{(2)}\right),
\end{equation}
where $\Mat{Z}_{vc}$ denotes the probability of assigning node $v$ to the class $c$; $\Mat{\hat{A}} = \Mat{\tilde{D}}^{-\frac{1}{2}}\Mat{\tilde{A}}\Mat{\tilde{D}}^{-\frac{1}{2}}$ is calculated by 
$\Mat{\tilde{A}} = \Mat{A} + \Mat{I}$ and the diagonal matrix $\Mat{\tilde{D}}$ with diagonal element $\Mat{\tilde{D}}_{ii} = \sum_{j}\Mat{\tilde{A}}_{ij}$ for $i=1,2,\cdots,n$; $\bm{\theta}=\left\{\Mat{W}^{(1)},\Mat{W}^{(2)}\right\}$ collects all the trainable parameters; $\sigma(\cdot)$ is an activation function (ReLU is used in this paper). In the framework of semi-supervised classification, the optimal parameter $\theta^*$ is learned by minimizing the cross-entropy loss on the labeled nodes $\mathcal{V}_{L}$, \ie,
\begin{align}
\label{gcntrain}
\min_{\bm{\theta}}\mathcal{L}_{train}\left(f_{\bm{\theta}}\left(\mathcal{G}\right)\right)=-\sum_{v\in \mathcal{V}_{L}}ln\Mat{Z}_{vc_v}.
\end{align}

\subsection{Problem Definition and Methodology}
\label{sec:prob_defi_method}
Different from previous works \citep{dai2018adversarial,zugner2018adversarial,zugner2019adversarial}, in this paper, we consider a new scenario: attacking a specific target node $v_0\in\mathcal{V}$ to change its prediction by injecting $n_{in}$ vicious nodes that are not on the original graph, denoted by $\mathcal{V}_{in}$ with $|\mathcal{V}_{in}|=n_{in}$ and $\mathcal{V}_{in}\cap\mathcal{V}=\emptyset$.
Formally, let $\mathcal{G}^\prime = \left(\Mat{A}^\prime,\Mat{X}^\prime\right)$ be the new graph after performing small perturbations on the original graph 
$\mathcal{G}$, then we have
\begin{equation}
\label{eq4} 
\Mat{A}^\prime = 
\left[
\begin{matrix}
\Mat{A} & \Mat{E}\\
\Mat{E}^\top & \Mat{O}
\end{matrix}
\right],\ 
\Mat{X}^\prime = \left[
\begin{matrix}
\Mat{X}\\
\Mat{X}_{in}
\end{matrix}
\right].
\end{equation}
Here $\Mat{E}\in \{0,1\}^{n\times n_{in}}$ denotes the relationship matrix between original nodes and vicious nodes. Specifically, $\Mat{E}_{ij}=1$ if the original node $v_i\in\mathcal{V}$ is connected to the vicious node $v_j\in\mathcal{V}_{in}$, and $\Mat{E}_{ij}=0$ otherwise. 
Symmetric matrix $\Mat{O}\in\{0,1\}^{n_{in}\times n_{in}}$ represents the relationships between vicious nodes in $\mathcal{V}_{in}$. 
The edge information denoted by $\Mat{E}$ and $\Mat{O}$ are called vicious edges in this paper.
$\Mat{X}_{in}\in \{0,1\}^{n_{in}\times d}$ is the feature matrix of vicious nodes. 
It is noteworthy that the perturbations can only be performed on $\Mat{E},\Mat{O}$ and $\Mat{X}_{in}$ while $\Mat{A}$ and $\Mat{X}$ remain unchanged.

Formally, the problem of adversarial attacks on graph $\mathcal{G}$ in the new scenario is typically formulated as the following bi-level optimization problem \footnote{The problem is formulated as bi-level optimization because the perturbed test input is also used in the training procedure and the model weight is dependent on perturbed test data.}:
\begin{align}
\label{eq0}
&\min_{\{\Mat{E},\Mat{O},\Mat{X}_{in}\}\in \Phi(\mathcal{G}^\prime)} \mathcal{L}_{atk}\left(f_{\bm{\theta}^*}\left(\mathcal{G}^\prime\right)\right) =\Mat{Z}^\prime_{v_0c_{v_0}} - \max_{c_{new}\neq c_{v_0}}\Mat{Z}^\prime_{v_0c_{new}}\\
&\ \ \ \ \ \ \ \ \ \mathrm{s.t.}\ \ \ \bm{\theta}^{*} = \arg\min_{\theta}\mathcal{L}_{train}\left(f_{\bm{\theta}}\left(\mathcal{G}^\prime\right)\right).\notag
\end{align}
where $\Mat{Z}'=f_{\bm{\theta}^*}(\mathcal{G}^\prime)$ denotes the prediction confidence scores for all classes on the perturbated graph $\mathcal{G}^{\prime}$, $\mathcal{L}_{atk}$ is the loss function during attack, $\Phi(\mathcal{G^{\prime}})$ is the constraints that $\Mat{E}$, $\Mat{O}$ and $\Mat{X}_{in}$ should meet on the perturbed graph $\mathcal{G}^{\prime}$. In the inner optimization problem, we get the optimal weights $\bm{\theta}^*$ of the model $f$(\ie, GCN) on the current perturbed graph $\mathcal{G}^\prime$ and in the outer optimization problem, we get the optimal perturbation(\ie, $\Mat{E}$, $\Mat{O}$ and $\Mat{X}_{in}$) on the current model $f_{\bm{\theta}^*}$. Subsequently, we will introduce the loss function on attack and constraint conditions in details.


\paragraph{{\bfseries Loss function of attacks.}}
The loss function of attacks $\mathcal{L}_{atk}$ aims to find a perturbed graph $\mathcal{G}^\prime$ that classifies the target node $v_0$ as $c_{new}$ and has maximal distance to the ground truth label $c_{v_0}$ in terms of log-probabilities/logits.
Thus, the smaller $\mathcal{L}_{atk}$ is, the worse the classification performs on the target node $v_0$. 
It is noteworthy that the surrogate model proposed in \citep{zugner2018adversarial} is typically used to generate perturbations instead of using Eq. \eqref{eq2} such that:
\begin{equation}
\label{eq5}
\Mat{Z} = f_{\bm{\theta}}(\mathcal{G}) = \hat{\Mat{A}}^2\Mat{X}\Mat{W}.
\end{equation}
The simplification is necessary for efficiency since it enables us to derive an approximate optimal solution (shown in section 4). In this sense, the loss function on attack turns to 
\begin{align}\label{loss_atk}
    \mathcal{L}_{atk}\left(f_{\bm{\theta}^{*}}\left(\mathcal{G}^\prime\right)\right)
    =\left[\hat{\Mat{A}^\prime}^2\Mat{X}^\prime\Mat{W}\right]_{v_0c_{v_0}} - \max_{c_{new}\neq c_{v_0}}\left[\hat{\Mat{A}^\prime}^2\Mat{X}^\prime\Mat{W}\right]_{v_0c_{new}}.
\end{align}

\paragraph{{\bfseries Constraint conditions.}} \label{sec:constraints}
There should be some constraint conditions to ensure unnoticeable perturbations, \ie, the definition of $\Phi(\mathcal{G}^\prime)$. 
One of the most widely used constraints for the adversarial attack is $\ell_0$-norm constraint. 
Specifically, the number of vicious edges by a budget $\Delta_e$ should be sparse, \ie,
\begin{equation}
\label{eq:constraint}
\|\Mat{E}\|_{0} + 0.5\|\Mat{O}\|_0 \leq \Delta_{e},
\end{equation}
where $\|\cdot \|_0$ denotes the $\ell_0$-norm of a matrix (\ie, the number of non-zero elements of a matrix).
Additionally, if we inject vicious nodes with strange pairs of features (\eg, mutually exclusive features), it will be easily detected. 
For this issue, the work in \citep{zugner2018adversarial} proposes a statistical test based on the co-occurrence graph of features to decide that a feature is unnoticeable if it occurs together with a node's original features. 
However, note that there are no original features for vicious nodes in our new scenario. Indeed, we can design the features arbitrarily. In this paper, we take a more practical solution by modifying the constraint such that the vicious nodes cannot import co-occurrence pairs of features that do not exist in the original graph. Formally:
\begin{equation}\label{eq:constraint_x}
\begin{aligned}
 \exists \  f_i,f_j \in \mathcal{F}, \ &\exists \  u_{in}\in \mathcal{V}_{in}, 
\  \left[\Mat{X}_{in}\right]_{u_{in},i} = 1 \wedge \left[\Mat{X}_{in}\right]_{u_{in},j} = 1\\
only \ if \ &\exists u \in \mathcal{V}, \Mat{X}_{u,i} = 1 \wedge \Mat{X}_{u,j} = 1
\end{aligned}
\end{equation}
This constraint means that if feature $i$ and feature $j$ occur together on the vicious node $u_{in}$, they must have occurred together in an original node $u$. In other words, no new co-occurrence pairs of features will be imported on the vicious nodes. Moreover, considering that vicious nodes with too few or too many features may be noticeable, we constrain the $\ell_0$-norm of vicious nodes to be equal to the mean of $\ell_0$-norm of original nodes, \ie, $\forall \ u_{in} \in \mathcal{V}_{in}, \|\left[\Mat{X}_{in}\right]_{u_{in}\cdot}\|_{0} = \|\Mat{X}\|_0/n$ where $\left[\Mat{X}_{in}\right]_{u_{in}\cdot}$ is the $u_{in}$-th row of matrix $\Mat{X}_{in}$ (\ie, the feature vector of vicious node $u_{in}$).

\section{The Proposed Framework}
It is a very challenging problem to solve the proposed optimization problem in Eq.~\eqref{eq0} due to the discreteness of the graph-structured data. In this section, we will first discuss how to adapt existing methods to the new attack scenario: injecting vicious nodes to perturb graphs. Moreover, considering the high computational complexity of existing methods, we propose a novel method (AFGSM) to speed up the computation and hence the developed method is scalable to large-scale graphs. Note that although we only focus on undirected graphs, all algorithms in this paper can be easily generalized to directed graphs.


\subsection{Adapting Existing Methods for the New Scenario}
In this paper, we consider three methods designed for the traditional scenario, including Nettack \citep{zugner2018adversarial}, FGSM \citep{zugner2018adversarial} and Meta-attack \citep{zugner2019adversarial}. To adapt them to the new attack scenario, we generate vicious nodes under the constraint conditions in Section \ref{sec:constraints} which is designed specifically for the new attack scenario instead of the constraints in \citep{zugner2018adversarial}.

There are two strategies to adapt existing methods to the new scenario. First, we can inject all vicious nodes to the graph, and then consider the vicious nodes as special original ones. We call this strategy as \textbf{one-time injection}. Second, we can inject vicious nodes one by one and once we inject a vicious node, we optimize the corresponding edges and features. We call this strategy \textbf{sequential injection}. 

\paragraph{\bfseries Nettack for the new scenario.}
Nettack\citep{zugner2018adversarial} addresses the bi-level optimization problem following a greedy strategy. Specifically, we initialize the vicious edges in the graph $\mathcal{G}^\prime$ with $\Mat{E}, \Mat{O}$, and set the initial $\Mat{X}_{in}$ by randomly sampling from the original graph $\mathcal{G}$. Then we apply Nettack on the graph $\mathcal{G}^\prime$ and constrain that perturbations are performed on $\Mat{E}, \Mat{O}$ and $\Mat{X}_{in}$. 
In each iteration, Nettack assigns a score for each potential edge that satisfies the constraints and choose the edge with the maximum score to flip. The main cost of Nettack is the calculations of the scores, thus it derives an incremental update method that can get the new scores from the old scores after each iteration in constant time. 

According to the analysis in \citep{zugner2018adversarial}, the time complexity of Nettack in terms of $n_{in}$ vicious nodes is $\mathcal{O}\left(\Delta_e\cdot n_{inj}\cdot(n\cdot Nei_{v_0}+fd)\right)$, where $Nei_{v_0}$ is the number of the one-order and second-order neighbors of the target node $v_0$, $f$ is the number of non-zero features in each vicious node's feature vector(\ie, $\|X\|_0/n$). Specifically, there are $\Delta_e$ iterations. In each iteration, $n$ scores for each vicious node should be calculated by considering all non-zero elements in $[\hat{\Mat{A}}^{\prime2}]_{v_0 \cdot}$ ($Nei_{v_0}$ edges average). And for each vicious node, once we select a feature ($f$ at most) for it, we need to check whether the candidate features ($d$ at most) of the node is co-occurring with it. 
Note that the strategies we adopt won't affect the computational complexity. 

\paragraph{\bfseries Meta-attack for the new scenario.}
Meta-attack\citep{zugner2019adversarial} utilizes the idea of meta-learning to optimize the perturbations generated on graphs.  
Note that Meta-attack is originally designed to attack the whole graph rather than attacking a specific target node $v_0$.
To apply Meta-attack to the new attack scenario, we replace the loss function $\mathcal{L}_{atk}$ with Eq.~\eqref{loss_atk} and update
the meta-gradients by
\begin{equation}
\begin{aligned}
\nabla_{\mathcal{G}^\prime}^{meta} &= \nabla_{\mathcal{G}^\prime}\mathcal{L}_{atk}(f_{\bm{\theta}^*}({\mathcal{G}^\prime})) \\
s.t.\ \ \bm{\theta}^*&= opt_{\bm{\theta}}(\mathcal{L}_{train}(f_{\bm{\theta}}({\mathcal{G}^\prime})))
\end{aligned}
\end{equation}
where $opt(\cdot)$ is a differentiable optimization procedure (\eg, gradient descent or its variants). 
In each iteration, Meta-attack picks the edge and the feature value with the maximum meta-gradient to flip. 

According to \citep{zugner2019adversarial}, the time complexity of Meta-attack for the new scenario is $\mathcal{O}\left(\Delta_e\left(Tn^2+n_{in}fd\right)\right)$, 
where $T$ is the number of iterations. For each iteration, the second-order gradient (Hessian matrix) should be calculated with a computational complexity of $\mathcal{O}\left(n^2\right)$. And for computation of features, it needs $\mathcal{O}\left(\Delta_en_{in}fd\right)$ just like Nettack. 

\paragraph{\bfseries FGSM for the new scenario.}
FGSM\citep{zugner2018adversarial} computes the gradients of $\mathcal{L}_{atk}$ \wrt edges and features and then it chooses the edge with the maximum revised gradient (\ie, multiply $-1$ if the edge exists) to flip and optimize the features with the signs of gradients. 

The time complexity of FGSM is $\mathcal{O}\left(\Delta_e\left(n^2+n_{in}fd\right)\right)$. For each iteration, the computation of gradients needs $\mathcal{O}\left(n^2\right)$. And like Nettack, the computation of features needs $\mathcal{O}\left(\Delta_en_{in}fd\right)$. 

\subsection{Approximate Fast Gradient Sign Method (AFGSM)}
Although existing methods can be adapted to the new attack scenario, their computational complexity is often too high to allow them to scale up in large-scale graphs in practice. To address this problem, we propose an efficient algorithm in this section, named Approximate Fast Gradient Sign Method (AFGSM) which injects vicious nodes one by one (\ie, the sequential injection strategy). 


Specifically, we inject vicious node $v_{in}$ with edge information $\Vec{e}_{in}\in\{0,1\}^{n}$ and feature vector $\Vec{x}_{in}\in \{0,1\}^{d}$ to attack the target node $v_0$ in graph $\mathcal{G}=\left(\Mat{A},\Mat{X}\right)$, and denote the new graph by $\mathcal{G}^\prime = (\Mat{A}^\prime, \Mat{X}^\prime)$, where
\begin{equation}
\label{eq8}
\Mat{A}^\prime = 
\left[
\begin{matrix}
\Mat{A} & \Vec{e}_{in}\\
\Vec{e}_{in}^\top & 0
\end{matrix}
\right]\in\{0,1\}^{(n+1)\times(n+1)},  
\Mat{X}^\prime = \left[
\begin{matrix}
\Mat{X}\\
\Vec{x}_{in}^\top
\end{matrix}
\right]\in\{0,1\}^{(n+1)\times d}.
\end{equation}
It is noteworthy that there are two possible values for the $v_0$-th component of edge information $\Vec{e}_{in}$, \ie, $\left[\Vec{e}_{in}\right]_{v_0}=0\ \text{or}\ 1$, where $[\cdot]_k$ refers to the $k$-th component of a vector.  
$\left[\Vec{e}_{in}\right]_{v_0}=1$ indicates that the vicious node $v_{in}$ is allowed to connect to the target node $v_0$ directly, namely \textbf{direct attack}; otherwise, we call it \textbf{indirect attack}, which is more difficult to be detected by intelligent defense algorithms since the vicious node is not connected to the target node directly.

Here we assume that in large-scale graphs, the changes of degrees of original nodes can be ignored after injecting only one vicious node, thus we can approximate the self-loop degree matrix $\tilde{\Mat{D}^\prime}$ as 


\begin{equation} \label{eq:approximation}
\tilde{\Mat{D}^\prime} \approx \left[
\begin{matrix}
\tilde{\Mat{D}}& \ \\
\ &\tilde{d}_{v_{in}}
\end{matrix}
\right]\in\mathbb{R}^{(n+1)\times(n+1)},
\end{equation}
where $\tilde{d}_{v_{in}} = d_{v_{in}}+1$; $d_{v_{in}}$ refers to the predefined degree of the vicious node $v_{in}$ (\ie, how many connections it will build with existing nodes). In this sense, the Laplacian matrix of graph $\mathcal{G}^\prime$ is calculated by
\begin{equation*} 
\hat{\Mat{A}^\prime} = \tilde{\Mat{D}^\prime}^{-\frac{1}{2}}(\Mat{A}^\prime + \Mat{I})\tilde{\Mat{D}^\prime}^{-\frac{1}{2}} 
\approx
\left[
\begin{matrix}
\hat{\Mat{A}}& \hat{\Vec{e}}_{in} \\
\hat{\Vec{e}}_{in}^\top & \tilde{d}_{v_{in}}^{-1}
\end{matrix}
\right]\in\mathbb{R}^{(n+1)\times(n+1)},
\end{equation*}
where $\hat{\Vec{e}}_{in} =
(\tilde{d}_{v_{in}})^{-\frac{1}{2}}\tilde{\Mat{D}}^{-\frac{1}{2}}\Vec{e}_{in}$.
As a result,  the probability of classification $\Mat{Z}^\prime$ after perturbation is derived as
\begin{equation} 
\label{eq14}
\begin{aligned}
\Mat{Z}^\prime 
= \hat{\Mat{A}'}^2\Mat{X}'\Mat{W} 
\approx
\left[\begin{matrix}
\hat{\Mat{A}}^2\Mat{X} + \hat{\Vec{e}}_{in}\hat{\Vec{e}}_{in}^\top\Mat{X} + \hat{\Mat{A}}\hat{\Vec{e}}_{in}\Vec{x}_{in}+\tilde{d}_{v_{in}}^{-1}\hat{\Vec{e}}_{in}\Vec{x}_{in} \\
\hat{\Vec{e}}_{in}^\top\hat{\Mat{A}}\Mat{X}+\tilde{d}_{v_{in}}^{-1}\hat{\Vec{e}}_{in}^\top\Mat{X}+\hat{\Vec{e}}_{in}^\top\hat{\Vec{e}}_{in}\Vec{x}_{in}+\tilde{d}_{v_{in}}^{-2}\Vec{x}_{in}
\end{matrix}\right]\Mat{W}.
\end{aligned}
\end{equation}
Specifically, the probability of classification for the target node $v_0$ turns to
{\small
\begin{align}
\label{Z'}
\Mat{Z}^\prime_{v_0j}
&\approx [\hat{\Mat{A}}^2\Mat{X}\Mat{W}]_{v_0j} + [\hat{\Vec{e}}_{in}]_{v_0}\hat{\Vec{e}}_{in}^\top\Mat{X}[\Mat{W}]_{\cdot j} + \left(\hat{\Vec{e}}_{in}^\top[\hat{\Mat{A}}]_{\cdot v_0} + \tilde{d}_{v_{in}}^{-1}[\hat{\Vec{e}}_{in}]_{v_0}\right)\Vec{x}_{in}\left[\Mat{W}\right]_{\cdot j}
\end{align}}for $j=1,2,\cdots,\left|\mathcal{C}\right|$, where  $[\cdot]_{i\cdot}$ and $[\cdot]_{\cdot j}$ denote the $i$-th row and $j$-th column of a matrix, respectively. $[\cdot]_{ij}$ refers to the $i$-th row and $j$-th component in a matrix.

Based on the approximation above, we observe that $\Mat{Z}^\prime_{v_0j}$ turns out to be linear with feature vector $\Vec{x}_{in}\in \{0,1\}^{d}$ and edge information $\Vec{e}_{in} \in \{0,1\}^{n}$.
Since the loss function $\mathcal{L}_{atk}$ formulated in Eq. \eqref{loss_atk} is also linear with $\Mat{Z}^\prime_{v_0j}$, we can obtain the optimal closed-form solutions of $\Vec{x}_{in}$ and $\Vec{e}_{in}$ which minimize loss function $\mathcal{L}_{atk}$ by their gradients.

\paragraph{{\bfseries An approximate closed-form solution of $\Vec{x}_{in}$.}}
From Eq. \eqref{Z'}, the output $\Mat{Z}^\prime$ is linear with respect to the variable $\Vec{x}_{in}$. Therefore, a closed-form solution of $\Vec{x}_{in}$ w.r.t. the optimization problem in Eq.~\eqref{eq0} can be obtained as follows
\begin{align} 
\label{eq13} 
\Vec{x}^*_{in} 
= - 0.5 sign\left(\frac{\partial \mathcal{L}_{atk}}{\partial \Vec{x}_{in}}\right) + 0.5,
\end{align}
where $sign(\cdot)$ is the element-wise function that takes the sign of a value. 
In other words, the features are set to 1 if the corresponding gradients in $\frac{\partial \mathcal{L}_{atk}}{\partial \Vec{x}_{in}}$ are negative and 0 otherwise.  
The gradients of loss function $\mathcal{L}_{atk}$ w.r.t. the variable $\Vec{x}_{in}$ can be calculated as follows
\begin{align}
\label{partialx}
\frac{\partial \mathcal{L}_{atk}}{\partial \Vec{x}_{in}} 
&\approx \tilde{d}_{v_{in}}^{-1}\left(\Vec{e}_{in}^\top\tilde{\Mat{D}}^{-\frac{1}{2}}[\hat{\Mat{A}}]_{\cdot v_0} + \tilde{d}_{v_{in}}^{-1}\tilde{d}_{v_{0}}^{-\frac{1}{2}}\left[\Vec{e}_{in}\right]_{v_0}\right)\left([\Mat{W}]_{\cdot c_{v_0}} -[\Mat{W}]_{\cdot c_{new}}\right), 
\end{align}
where $\tilde{d}_{v_{in}}^{-1}(\Vec{e}_{in}^\top\tilde{\Mat{D}}^{-\frac{1}{2}}[\hat{\Mat{A}}]_{\cdot v_0} +\tilde{d}_{v_{in}}^{-1}\tilde{d}_{v_{0}}^{-\frac{1}{2}}\left[\Vec{e}_{in}\right]_{v_0})$ is always positive, and thus can be ignored since we only care about the signs of the gradients. 

\paragraph{{\bfseries An approximate closed-form solution of $\Vec{e}_{in}$.}}
Without loss of generality, we assume that the vicious node $v_{in}$ connects to a fixed number of nodes in the original graph $\mathcal{G}$, and therefore $d_{v_{in}}=\|\Vec{e}_{in}\|_0$ holds.  
Since the output $\Mat{Z}^\prime$ is linear w.r.t. the variable $\vec{e}_{in}$, we achieve a closed-form solution of $\Vec{e}_{in}$ with constraint in Eq.~\eqref{eq:constraint}, \ie,
\begin{equation}
\label{optimal_e}
    \Vec{e}_{in}^* = -0.5sign\left(\frac{\partial \mathcal{L}_{atk}}{\partial \Vec{e}_{in}} - g_{d_{in}}\Vec{1}\right)+0.5,
\end{equation}
where $g_{d_{in}}$ is the $d_{in}$-th smallest element of $\frac{\partial \mathcal{L}_{atk}}{\partial \Vec{e}_{in}}$. 
In other words, the $i$-th component of $\Vec{e}_{in}^*$ is set to 1 if the corresponding gradients are less than the $d_{in}$-th smallest gradient in $\frac{\partial \mathcal{L}_{atk}}{\partial \Vec{e}_{in}}$. 
The gradients of loss function $\mathcal{L}_{atk}$ w.r.t. the variable $\Vec{e}_{in}$ is derived as follows
{\small
{\begin{align}
\frac{\partial \mathcal{L}_{atk}}{\partial \Vec{e}_{in}} 
&\approx \left(\tilde{d}_{v_{in}}\tilde{d}_{v_0}\right)^{-\frac{1}{2}}\left(\tilde{d}_{v_{in}}^{-\frac{1}{2}}\left[\Vec{e}_{in}\right]_{v_0}\tilde{\Mat{D}}^{-\frac{1}{2}}\Mat{X}+\tilde{\Mat{D}}^{-1}[\tilde{\Mat{A}}]_{\cdot v_0}\Vec{x}_{in}\right)\left([\Mat{W}]_{\cdot c_{v_0}} -[\Mat{W}]_{\cdot c_{new}}\right), \label{partiale}
\end{align}}}
where $(\tilde{d}_{v_{in}}\tilde{d}_{v_0})^{-\frac{1}{2}}$ is always positive, and thus can be ignored.

\begin{algorithm}[t]
	\caption{The proposed AFGSM\label{alg3} algorithm.}
	\begin{algorithmic}[1]
		\Require  Graph $\mathcal{G} = (\Mat{A},\Mat{X})$, the target node $v_0$, the number of vicious nodes $n_{in}$, the budget of edge perturbations $\Delta_e$;
		\State Train the surrogate model on the original graph $\mathcal{G}$ and get its weight matrix $\Mat{W}$;
		\State Randomly assign the degrees of vicious nodes $d_{v_{in}}^{(0)}, d_{v_{in}}^{(1)},\cdots, d_{v_{in}}^{(n_{in}-1)}$ to satisfy the budget constraint $\sum_{k=0}^{n_{in}-1}d_{v_{in}}^{(k)}=\Delta_e$;
		\State $\mathcal{G}'^{(0)}\leftarrow \mathcal{G}, \Mat{A}^{(0)}\leftarrow \Mat{A}, \Mat{X}^{(0)}\leftarrow \Mat{X}$;
		\State Calculate $\Vec{x}_{in}^*$ according to Eq. \eqref{eq13};
		\For{$t=0,\cdots,n_{in}-1$}
		\State Initialize $\Vec{e}_{in}^{(t)}=\Vec{0}, \Vec{x}_{in}^{(t)}=\Vec{0}$;
		\State $\Vec{x}_{in}^{(t)}\leftarrow$ Sample $\|X\|_{0}/n$ features from $\Vec{x}_{in}^*$ under the constraint condition $\Phi(\mathcal{G})$;
	    \State $\Vec{e}^{(t)}\leftarrow$ Calculate $\Vec{e}_{in}^*$ according to Eq. \eqref{optimal_e};
	    \State $\mathcal{G}^{\prime(t+1)}=\left(\Mat{A}^{(t+1)}, \Mat{X}^{(t+1)}\right)$ $\leftarrow$ Update $\Mat{A}^{(t)}$ and $\Mat{X}^{(t)}$ according to Eq. \eqref{eq8};
		\EndFor
		\State \textbf{return} $\mathcal{G}^{\prime(n_{in})}=\left(\Mat{A}^{(n_{in})},\Mat{X}^{(n_{in})}\right)$; 
	\end{algorithmic}		 
\end{algorithm}
\begin{algorithm}[t]
	\caption{The proposed AFGSM-ada\label{alg4} algorotihm.}
	\begin{algorithmic}[1]
		\Require  Graph $\mathcal{G} = (\Mat{A},\Mat{X})$, the target node $v_0$, the number of vicious nodes $n_{in}$, the budget of edge perturbations $\Delta_e$;
		\State Randomly assign the degrees of vicious nodes $d_{v_{in}}^{(0)}, d_{v_{in}}^{(1)},\cdots, d_{v_{in}}^{(n_{in}-1)}$ to satisfy the budget constraint $\sum_{k=0}^{n_{in}-1}d_{v_{in}}^{(k)}=\Delta_e$;
		\State $\mathcal{G}'^{(0)}\leftarrow \mathcal{G}, \Mat{A}^{(0)}\leftarrow \Mat{A}, \Mat{X}^{(0)}\leftarrow \Mat{X}$;
		\For{$t=0,\cdots,n_{in}-1$}
		\State Train the surrogate model on the perturbed graph $\mathcal{G}^{\prime(t)}$ and get its weight matrix $\Mat{W}$;
		\State Calculate $\Vec{x}_{in}^*$ according to Eq. \eqref{eq13};
		\State Initialize $\Vec{e}_{in}^{(t)}=\Vec{0}, \Vec{x}_{in}^{(t)}=\Vec{0}$;
		\State $\Vec{x}_{in}^{(t)}\leftarrow$ Sample $\|X\|_{0}/n$ features from $\Vec{x}_{in}^*$ under the constraint condition $\Phi(\mathcal{G})$;
	    \State $\Vec{e}^{(t)}\leftarrow$ Calculate $\Vec{e}_{in}^*$ according to Eq. \eqref{optimal_e};
	    \State $\mathcal{G}^{\prime(t+1)}=\left(\Mat{A}^{(t+1)}, \Mat{X}^{(t+1)}\right)$ $\leftarrow$ Update $\Mat{A}^{(t)}$ and $\Mat{X}^{(t)}$ according to Eq. \eqref{eq8};
		\EndFor
		\State \textbf{return} $\mathcal{G}^{\prime(n_{in})}=(\Mat{A}^{(n_{in})}$, $\Mat{X}^{(n_{in})})$;
	\end{algorithmic}		 
\end{algorithm}


We summarize the procedure of the proposed AFGSM in Algorithm \ref{alg3}.
Note that the approximate closed-form solution of feature vector $\Vec{x}^*_{in}$ is calculated for once as it only depends on the weights $\Mat{W}$. However, the approximate closed-form solution of edge information $\Vec{e}_{in}^*$ relies on variables $\Mat{A}$ and $\Mat{X}$ and thus should be updated as the vicious nodes are injected one by one in a sequential manner. 

Note that Algorithm \ref{alg3} treats the model weight matrix $\Mat{W}$ as static. Alternatively, we also develop an adaptive version of AFGSM, namely \emph{AFGSM-ada} by re-training the surrogate model once we inject a vicious node using the AFGSM. We summarize the procedure of \emph{AFGSM-ada} in Alghritm \ref{alg4}. Similarly, we also develop the adaptive version of Nettack and FGSM correspondingly, namely \emph{Nettack-ada} and \emph{FGSM-ada}. As for Meta-attack, it updates the model weights dynamically, so there is no need to develop its variant.

\paragraph{{\bfseries Complexity of AFGSM.}}
The time complexity of the proposed AFGSM algorithm is $\mathcal{O}\left(n_{in}\left( n+fd\right)\right)$. This is because that the gradients in Eq. \eqref{partialx} and Eq. \eqref{partiale} is concise enough such that they can be calculated by several vectors and no matrix multiplication is involved. And for each vicious node, the computation of features needs $\mathcal{O}\left(fd\right)$ just like analyzed above. 

\subsection{Comparison of Complexity}
\label{sec:complexity}
Comparing the time complexity of different methods, we observe that 
\begin{align*}
\text{AFGSM:} &\mathcal{O}\left(n_{in}\left( n+fd\right)\right) < \text{Nettack:} \mathcal{O}\left(\Delta_e\cdot n_{inj}\cdot\left(n\cdot Nei_{v_0}+fd\right)\right) <\\ \text{FGSM:}  &\mathcal{O}\left(\Delta_en^2+n_{in}fd\right) < 
\text{Meta-attack:} \mathcal{O}\left(\Delta_eTn^2+n_{in}fd\right)
\end{align*}

As a result, the proposed AFGSM algorithm is the most efficient one in terms of time complexity and then followed by Nettack, FGSM and Meta-attack. 
Experimental results also substantiate the conclusion in the next section.

\section{Experiments}
\label{sec:evaluation}

\begin{table}[t]
	\centering
	\caption{The detailed statistics of the used datasets. $N_{LCC}$ and $E_{LCC}$ are the numbers of nodes and edges in the largest connected component, $d$ is the dimension of features and $C$ is the number of classes.}
	\label{table1}
	\resizebox{\textwidth}{!}{
	\begin{tabular}{l|r|r|r|r|l}
		\toprule
		Dataset&$N_{LCC}$&$E_{LCC}$&$d$&$C$&frequency of classes\\
		\midrule
		Citeseer&2,110&3,668&3,703&6&532, 463, 388, 308, 304, 115\\
		Cora&2,485&5,069&1,433&7&726, 406, 379, 344, 285, 214, 131\\
		DBLP&16,766&44,422&2,476&4&6935, 6532, 1777, 1522\\
		Pubmed&19,717&44,324&500&3&7875, 7739, 4103\\
		Reddit&149,177&5,215,380&602&41&28163, 15163, 13963, 13065, 12742, ... \tablefootnote{The frequency of classes are: 28163, 15163, 13963, 13065, 12742, 12041, 11149, 10239,  7915, 5863,  5087, 5048,  4937,  4898,  4849,  4668,  4547,  4212, 4188, 4184,  4161,  4040,  3930,  3588,  3538,  3422,  3279, 2970,  2960,  2792,  2687, 2630,  2304,  2232,  2115, 1696, 1645, 1588, 1554, 991, 328} \\
		\bottomrule
	\end{tabular}}
\end{table}

In this section, we evaluate the effectiveness and efficiency of our attack in the new attack scenario. We first provide the experimental setup (Section \ref{sec:exp_setup}). Then we show the results of adapting existing methods to the new attack scenario following two different strategies: one-time injection and sequential injection (Section \ref{sec: exist methods}). Next, we explore the performance of our methods on large graphs and analyze the time cost (Section \ref{sec:AFGSM}). Finally, we consider two stricter scenarios where attackers have limited capability and show the performance of the AFGSM method in these restricted cases (Section \ref{sec:strict limits}). 

\subsection{Experimental Setup} 
\label{sec:exp_setup}
We conduct our experiments on five well-known public datasets:  Citeseer~\citep{sen2008collective}, Cora~\citep{mccallum2000automating}, Pubmed~\citep{sen2008collective}, DBLP~\citep{zhang2019attributed} and Reddit \citep{hamilton2017inductive}. The first four are citation networks where nodes are documents, edges are citation links and features are selected as the words in the document after filtering out the stop words. And the last one is a post-to-post graph where nodes are posts and edges denote these posts are from the same user. Due to the high cost of training models (\eg, GCN and Deepwalk) on the original Reddit graph(around $230k$ nodes), we randomly sample a subgraph with nearly $150K$ nodes. The detailed statistics of these datasets are shown in Table \ref{table1}. Following the same attack setting in~\citep{zugner2018adversarial}, we only consider the largest connected component for convenience.  

In the experiments, we split the datasets into training set (10\%), validation set (10\%), and test set (80\%). Note that in practice, attackers rarely can manipulate the training data. Therefore, we only inject vicious nodes to the test set (without labels). We first train a surrogate model on the training set, and then among all nodes that are correctly classified, we select (i) the 10 nodes with the highest margin of classification, (ii) the 10 nodes with the lowest margin of classification, (iii) 30 nodes selected randomly as our target nodes to be attacked. Since we focus on transductive classification in this paper, the model is then retrained on the mixture of clean training and perturbed test data. For each target node, we repeat the retraining process 5 times with different random seeds to stabilize the results and the average performance is reported. 

\begin{table*}[t]
	\centering

	\setlength{\tabcolsep}{2pt}
	\caption{Accuracy of victim learning models against different attacks with one-time injection strategy w.r.t. different number of initial connections between vicious nodes and target node. \emph{Clean} denotes the model without any attacks. \emph{Random} denotes the model in which the vicious nodes connect to existing nodes randomly and their features are also sampled randomly according to those of existing nodes.}\label{tab:diff_initial_value}
    \resizebox{\textwidth}{!}{
	\begin{tabular}{lcccccccccccc}
		\toprule
		\multirow{2}{*}{Method}&
		\multicolumn{3}{c}{Citeseer}&\multicolumn{3}{c}{Cora}\cr
		\cmidrule(lr){2-4} \cmidrule(lr){5-7}
	    &GCN&GAT&Deepwalk&GCN&GAT&Deepwalk\cr
		\midrule
		Clean&$0.892\pm0.010$&$0.804\pm0.008$& $0.736\pm0.054$&$0.928\pm0.016$&$0.788\pm0.027$&$0.840\pm0.028$\cr
		Random &$0.772\pm0.020$&$0.708\pm0.016$&$0.648\pm0.041$&$0.868\pm0.029$&$0.700\pm0.022$&$0.728\pm0.063$\cr
		Nettack-0\%&$0.480\pm0.025$&$0.460\pm0.022$&$0.652\pm0.041$&$0.424\pm0.008$&$0.516\pm0.023$&$0.744\pm0.029$\cr
		Nettack-50\%&$0.336\pm0.029$&$0.304\pm0.015$&$0.592\pm0.081$&$0.352\pm0.016$&$0.428\pm0.016$&$0.656\pm0.064$\cr
		Nettack-100\%&$0.248\pm0.010$&$0.244\pm0.015$&$0.604\pm0.102$&$0.324\pm0.023$&$0.356\pm0.022$&$0.651\pm0.032$\cr
		Meta-attack-0\%&$0.148\pm0.020$&$0.156\pm0.015$&$0.460\pm0.046$&$0.204\pm0.015$&$0.272\pm0.020$&$0.484\pm0.061$\cr
		Meta-attack-50\%&$0.112\pm0.027$&$\bm{0.140\pm0.028}$&$0.412\pm0.059$&$0.224\pm0.039$&$0.232\pm0.010$&$0.520\pm0.073$\cr
		Meta-attack-100\%&$\bm{0.104\pm0.008}$&$0.164\pm0.015$&$\bm{0.352\pm0.081}$&$\bm{0.188\pm0.020}$&$\bm{0.196\pm0.023}$&$\bm{0.452\pm0.047}$\cr
		FGSM-0\%&$0.208\pm0.020$&$0.300\pm0.033$&$0.524\pm0.041$&$0.336\pm0.015$&$0.436\pm0.057$&$0.596\pm0.085$\cr
		FGSM-50\%&$0.200\pm0.013$&$0.216\pm0.023$&$0.504\pm0.034$&$0.260\pm0.013$&$0.304\pm0.015$&$0.536\pm0.048$\cr
		FGSM-100\%&$0.216\pm0.023$&$0.216\pm0.150$&$0.504\pm0.066$&$0.292\pm0.010$&$0.252\pm0.016$&$0.488\pm0.032$\cr
		\bottomrule
	\end{tabular}
	}
\end{table*}
\subsection{Adapting Existing Methods to the New Scenario}\label{sec: exist methods}
As mentioned in the previous section, the node injection process can be performed in two ways: inject nodes all at once (one-time injection) and inject nodes sequentially (sequential injection). In this section, we compare the performance of different attacks with these two different node injection strategies.

\paragraph{\bfseries{One-time injection.}}
The one-time injection can be roughly treated as the special case of the attack scenario considered in \citep{zugner2018adversarial} as nodes are added in advance and perturbations are generated using existing attacks\footnote{There is a difference in the constraint for feature perturbations. As explained in Section \ref{sec:prob_defi_method}, we do not have specific feature constraints (however, we do not allow the occurrence of pairs of features that do not exist in the original nodes) for the vicious nodes while in the original scenario, the number of feature perturbations cannot exceed a certain threshold.}. One-time injection proceeds by first connecting a fixed number of vicious nodes to the target node directly and then treat the newly added nodes as existing nodes and apply the attacks proposed in \citep{zugner2018adversarial,zugner2019adversarial}. We choose to connect the vicious nodes directly because connections to the target node usually lead to better attack performance~\citep{zugner2018adversarial}. However, we do not know the optimal number of vicious nodes that should be connected and testing all combinations of vicious nodes is not practical. Therefore, we randomly connect 0\%, 50\% and 100\% of the vicious nodes to the target node. Although this simple approach cannot cover all the cases, as shown below, connecting all vicious nodes (to the target node) gives the best attack performance. 

The results of different attacks under different number of initial connections (to the target node) are shown in Table \ref{tab:diff_initial_value}. Note that we enforce the same perturbation budget (10 nodes and 20 edges) for all methods in Table \ref{tab:diff_initial_value} for a fair comparison.  First, we observe that all attacks get the best performance when we connect 100\% of vicious nodes to the target node, which is also consistent with our intuition and the results in~\citep{zugner2018adversarial} that (more) connections with target node gives better attack results. Second, we find that Meta-attack performs the best in the one-time injection. The reason is that Meta-attack generates the perturbation on edges and features utilizing the second-order gradients which can provide more information by considering the model weights dynamically. However, the complexity of Meta-attack is often very high because of the higher-order gradients and hence, cannot scale to large graphs such as DBLP and Pumbed used in the paper. Third, interestingly, we observe that FGSM performs better than Nettack in the new scenario which is quite different from the results in \citep{zugner2018adversarial}. We hypothesize that it may be due to the limited search space of Nettack in the new scenario. More specifically, the initial values of $\Mat{E}$, $\Mat{O}$ are extremely sparse (\ie, only a small number of connected edges) compared to the number of existing nodes and edges in the graph, even if we connect 100\% of vicious nodes to the target node initially. Therefore, the search space for Nettack is severely limited. 

\paragraph{\bfseries{Sequential injection.}}
Next, we explore the performance of different attacks using the sequential injection strategy. Following the same setting in the one-time injection, we still connect the vicious nodes to the target node, but in a sequential manner. The results are shown in Table~\ref{tab:one-time injection}. 

We find that, in the sequential addition scenario, FGSM performs better while Meta-attack performs worse compared to their counterpart in the one-time injection scenario. For example, on the Cora dataset, FGSM-one-time only lowers the GCN accuracy from 92.8\% to 29.2\% while FGSM-sequential lowers the accuracy to 25.6\%. Differently, still on the Cora dataset and with the GCN model, Meta-attack-one-time can lower the accuracy from 92.8\% to 18.8\% while Meta-attack-sequential can only lower it to 24.8\%. As for Nettack, there is no significant difference in the performance by following different node injection strategies. Nettack-one-time can lower the accuracy on Cora to 32.4\% while Nettack-sequential lowers it to 31.6\%. For GCN on the Citeseer, Nettack-one-time lowers the accuracy to 24.8\% while Nettack-sequential lowers it to 25.2\%, which has a negligible difference.
We hypothesize that different performance of Meta-attack and FGSM under the two injection strategies is related to the search space identified by the node injection strategies and the nature of the attacks. 

The main difference between the two node injection strategies is the degree of freedom in their valid search space. For the sequential injection, attacks are limited to manipulate a limited number of edges for each vicious node as we have a constraint on the degree of each vicious node. Therefore, for attacks that utilize limited information (e.g., first-order gradient for FGSM), this limitation helps to prevent attacks from manipulating too many sub-optimal edges for a single node and hence avoids getting stuck into bad solutions. In contrast, for the one-time injection, attacks are only constrained by the total number of vicious edges and thus have higher freedom when perturbing edges. For attacks with limited information, a higher degree of freedom can easily lead to suboptimal solutions while with more information (e.g., second-order derivative for Meta-attack), a higher degree of freedom leads to better attack results. For Nettack, it still suffers from the limited search space with sequential injection (similar to the analysis in one-time injection) and hence, doesn't show significant difference under different node injection strategies. Moreover, we observe that FGSM with sequential injection performs relatively closely to the costly Meta-attack with the one-time injection and the gap can be further reduced by adaptively retraining the model during attack process and more details can be found in Table~\ref{tab:ada}. However, in comparison to the complicated second order in Meta-attack, the first-order gradient in FGSM can be easily approximated and hence more efficient \emph{AFGSM} is proposed in this paper.    

\vspace{-1dd}
For experiments presented in the rest part of the paper, when choosing the baseline methods, for each method, we select the best one under the two node injection strategies. Specifically, we select FGSM-sequential, Meta-attack-one-time, and Nettack-one-time. For convenience, we still denote them as FGSM, Meta-attack and Nettack, respectively.

\begin{table*}[t]
	\centering

	\setlength{\tabcolsep}{1pt}
	\caption{Accuracy of victim learning models against different attacks with two different strategies. {Attacks with a postfix of \emph{one-time} are attacks with one-time node injection strategy. Attacks with a postfix \emph{sequential} are attacks with sequential node injection strategy.}}\label{tab:one-time injection}

	\resizebox{\textwidth}{!}{
	\begin{tabular}{lcccccccccccc}
		\toprule
		\multirow{2}{*}{Method}&
		\multicolumn{3}{c}{Citeseer}&\multicolumn{3}{c}{Cora}\cr
		\cmidrule(lr){2-4} \cmidrule(lr){5-7}
	    &GCN&GAT&Deepwalk&GCN&GAT&Deepwalk\cr
		\midrule
		Clean&$0.892\pm0.010$&$0.804\pm0.008$& $0.736\pm0.054$&$0.928\pm0.016$&$0.788\pm0.027$&$0.840\pm0.028$\cr
		Nettack-one-time&$0.248\pm0.010$&$0.244\pm0.015$&$0.604\pm0.102$&$0.324\pm0.023$&$0.356\pm0.022$&$0.651\pm0.032$\cr
		Nettack-sequential&$0.252\pm0.024$&$0.228\pm0.010$&$0.644\pm0.066$&$0.316\pm0.015$&$0.356\pm0.020$&$0.784\pm0.065$\cr
		Meta-attack-one-time &$\bm{0.104\pm0.008}$&$0.164\pm0.015$&$\bm{0.352\pm0.081}$&$\bm{0.188\pm0.020}$&$\bm{0.196\pm0.023}$&$0.452\pm0.047$\cr
		Meta-attack-sequential&$0.120\pm0.012$&$0.228\pm0.016$&$0.392\pm0.071$&$0.248\pm0.024$&$0.240\pm0.013$&$\bm{0.420\pm0.046}$\cr
		FGSM-one-time&$0.216\pm0.023$&$0.216\pm0.150$&$0.504\pm0.066$&$0.292\pm0.010$&$0.252\pm0.016$&$0.488\pm0.032$\cr
		FGSM-sequential&$0.156\pm0.020$&$\bm{0.144\pm0.015}$&$0.408\pm0.061$&$0.256\pm0.008$&$0.316\pm0.008$&$0.456\pm0.023$\cr
		\bottomrule
	\end{tabular}
	}
\end{table*}

\subsection{Attack by AFGSM and its adaptive variant.}\label{sec:AFGSM} 
We conduct experiments on small and large-scale graphs to show the effectiveness and efficiency of our attack.
For each attack in this section, we additionally consider an adaptive variant of the attack, which trains the model dynamically during the attack process. Therefore, both AFGSM and FGSM have two attack forms: one with fixed model during the attack and one with dynamically retrained model during the attack. Meta-attack does not have its adaptive variant because the original attack already retrains the model during the attack. We denote the attack with adaptive training by adding a postfix ``ada''.

\begin{figure}[t] 
	\hspace{-0.5cm}
	\centering
	\includegraphics[width=1.0\columnwidth]{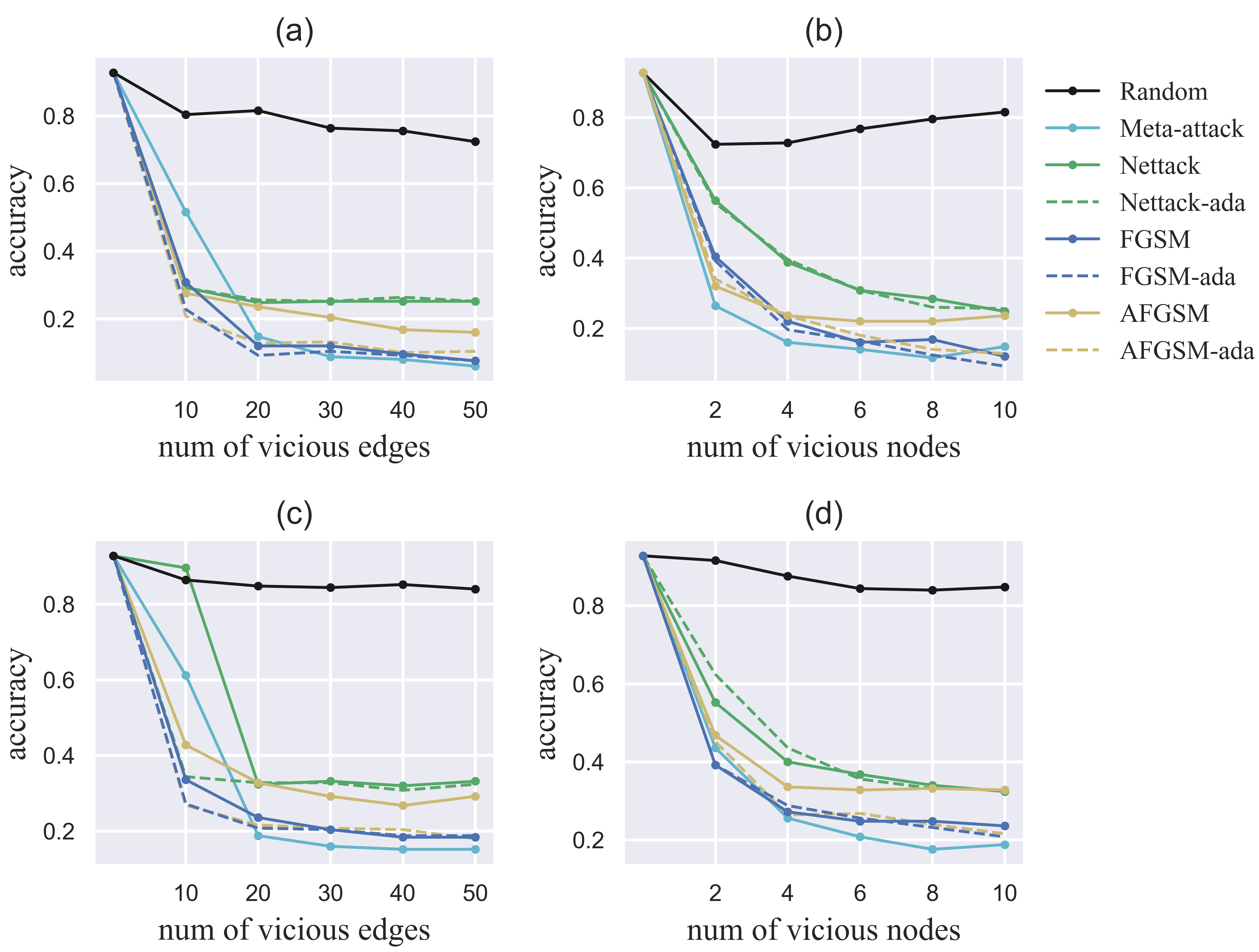}
	\caption{Accuracy on GCN with different numbers of vicious nodes and edges. (a) and (c) denote the performance of GCN with different numbers of vicious edges and 10 nodes on Citeseer and Cora, respectively. (b) and (d) denote the performance of GCN with different numbers of vicious nodes and 10 nodes on Citeseer and Cora, respectively.}\label{fig:diff_edges_nodes}
\end{figure}

\paragraph{\bfseries{On small graphs.}} First, we conduct experiments on small graphs (\eg, Citeseer and Cora). The results are shown in Table~\ref{tab:ada}. We find that adaptively training models during the attack process helps FGSM and AFGSM achieve better performance. For example, AFGSM for GCN on the Citeseer dataset only lowers the accuracy from 89.2\% to 22.4\% while AFGSM-ada lowers it to 10.4\%. Although Meta-attack still performs the best among all methods, FGSM-ada and AFGSM-ada get a pretty close performance to Meta-attack. AFGSM performs similar to FGSM because our approximation technique preserves the attack effectiveness while improves the efficiency significantly. To show that the good performance of AFGSM and AFGSM-ada does not depend on a specific set of hyperparameters, we also compare all the attacks using different constraint budgets. The results are shown in Figure \ref{fig:diff_edges_nodes}. We can easily find that under different sets of attack hyperparameters, AFGSM is still very effective and AFGSM-ada performs close to the best performing Meta-attack. We emphasize that AFGSM achieves the comparably good performance with much lower computational complexity. 

\begin{table*}[t]
	\centering

	\setlength{\tabcolsep}{2pt}
	\caption{Accuracy of victim learning models against different attacks and adaptive variants. Attacks with a postfix of \emph{ada} are adaptive attacks with victim learning models retrained during the attack process.}\label{tab:ada}
    \resizebox{\textwidth}{!}{
	\scalebox{1}{
	\begin{tabular}{lcccccccccccc}
		\toprule
		\multirow{2}{*}{Method}&
		\multicolumn{3}{c}{Citeseer}&\multicolumn{3}{c}{Cora}\cr
		\cmidrule(lr){2-4} \cmidrule(lr){5-7}
	    &GCN&GAT&Deepwalk&GCN&GAT&Deepwalk\cr
		\midrule
		Clean&$0.892\pm0.010$&$0.804\pm0.008$& $0.736\pm0.054$&$0.928\pm0.016$&$0.788\pm0.027$&$0.840\pm0.028$\cr
		Random &$0.772\pm0.020$&$0.708\pm0.016$&$0.648\pm0.041$&$0.868\pm0.029$&$0.700\pm0.022$&$0.728\pm0.063$\cr
		Nettack&$0.248\pm0.010$&$0.244\pm0.015$&$0.604\pm0.102$&$0.324\pm0.023$&$0.356\pm0.022$&$0.651\pm0.032$&\cr
		Nettack-ada&$0.256\pm0.008$&$0.224\pm0.023$&$0.556\pm0.069$&$0.304\pm0.023$&$0.304\pm0.019$&$0.656\pm0.064$\cr
		Meta-attack&$\bm{0.104\pm0.008}$&$0.164\pm0.015$&$\bm{0.352\pm0.081}$&$\bm{0.188\pm0.020}$&$\bm{0.196\pm0.023}$&$\bm{0.452\pm0.047}$\cr
		FGSM&$0.156\pm0.020$&$0.144\pm0.015$&$0.408\pm0.061$&$0.256\pm0.008$&$0.316\pm0.008$&$0.456\pm0.023$\cr
		FGSM-ada&$0.112\pm0.010$&$0.136\pm0.015$&$0.444\pm0.057$&$0.208\pm0.020$&$0.364\pm0.015$&$0.472\pm0.056$\cr
		\midrule
		AFGSM &0$.224\pm0.023$&$0.192\pm0.016$&$0.532\pm0.060$&$0.304\pm0.015$&$0.404\pm0.041$&$0.580\pm0.051$\cr
        AFGSM-ada&$0.104\pm0.027$&$\bm{0.128\pm0.020}$&$0.484\pm0.055$&$0.212\pm0.008$&$0.388\pm0.036$&$0.588\pm0.056$\cr
		\bottomrule
	\end{tabular}}
	}
\end{table*}

\paragraph{\bfseries{On large graphs.}}\label{sec:time cost}
To show the scalability of our algorithm, we first conduct experiments on large DBLP and Pubmed datasets (still with 10 vicious nodes and 20 edges). For graphs with $10K+$ nodes, we can still obtain results for Nettack, FGSM, and AFGSM and their adaptive variants. Meta-attack cannot scale to these graphs and hence, we do not include the results of Meta-attack. Details of the experiments can be found in Table \ref{tab:lager_graphs}. We observe similar results as of small graphs: AFGSM performs closely to FGSM, and their adaptive variants provide better performance. One exception happens for Deepwalk on DBLP, which might be because of the poor transferability of the adaptive attack on GCN to Deepwalk (attacks on GAT and Deepwalk are all transferred from the attacks on GCN). Among all methods, FGSM-ada performs the best and AFGSM-ada is close to FGSM-ada. However, due to high complexity, FGSM or FGSM-ada cannot scale to graphs with more than $30K$ nodes (tested on our machine\footnote{We record the actual run time on the same machine with configuration: CPU (i9-7900X, 3.30GHz), 128GB RAM.} using sampled subgraphs from Reddit).

We further test the performance of our attack on larger graphs, where none of the baseline attacks can scale. We construct the large graph by subsampling from the Reddit dataset with $150K$ nodes. Note that we only perturb edges because the features of Reddit are preprocessed and it is impractical to directly manipulate the preprocessed features. Considering that the results of target nodes with higher degrees are harder to be mislead~\citep{zugner2018adversarial}, we attack the target nodes with $d_{v_0}/2$ vicious nodes and $d_{v_0}$ edges. The results are shown in Table~\ref{tab:reddit}. Our AFGSM can still significantly reduce the accuracy on the GCN model. The transferability of the attack on GCN to Deepwalk is relatively low and we leave it as future work to improve the transferability of AFGSM on extremely large graphs. We also note that, when evaluating our attack on extremely large graphs, the main bottleneck will be in the model training instead of the attack process. Therefore, as long as there are efficient methods to train models on extremely large graphs, our attack can always scale and (highly probably) work.

\paragraph{\bfseries{Time cost analysis.}}

To compare the computational cost of different methods, we conduct experiments (on the same machine mentioned above) on subgraphs of DBLP with different sizes (\ie, varying from $1K$ to $15K$ nodes with a step size of $1K$ nodes). The results of time cost are shown in Figure~\ref{fig:time_cost}. It is obvious that AFGSM is much more efficient than other baseline attacks, and the order of time cost aligns well with the complexity analysis in Section \ref{sec:complexity} (\ie, AFGSM is the most efficient and followed by Nettack, FGSM, and Meta-attack). Furthermore, the execution time of AFGSM remains relatively stable when the size of the graph grows, while the run time of other attacks grow much faster when the graph size increases, especially for Meta-attack (time cost is 3,000 times higher than AFGSM and cannot scale to the DBLP dataset with more than $13K$ nodes). 
\begin{table*}[t]
	\centering
	\setlength{\tabcolsep}{2pt}
	\caption{Accuracy of victim learning models against different attacks on large graphs.}\label{tab:lager_graphs}
    \resizebox{\textwidth}{!}{
	
	\begin{tabular}{lcccccccccccc}
		\toprule
		\multirow{2}{*}{Method}&
		\multicolumn{3}{c}{DBLP}&\multicolumn{3}{c}{Pubmed}\cr
		\cmidrule(lr){2-4} \cmidrule(lr){5-7}
		&GCN&GAT&Deepwalk&GCN&GAT&Deepwalk\cr
		\midrule
		Clean&$0.880\pm0.025$&$0.808\pm0.010$&$0.856\pm0.023$&$0.904\pm0.022$&$0.848\pm0.020$&$0.832\pm0.020$\cr
		Random &$0.712\pm0.020$&$0.724\pm0.015$&$0.832\pm0.047$&$0.848\pm0.016$&$0.796\pm0.023$&$0.792\pm0.032$\cr
		Nettack&$0.268\pm0.016$&\bm{$0.420\pm0.021$}&$0.712\pm0.035$&$0.152\pm0.027$&$0.252\pm0.010$&$0.796\pm0.046$\cr
		Nettack-ada&$0.272\pm0.016$&$0.436\pm0.019$&$0.808\pm0.016$&$0.156\pm0.023$&$0.240\pm0.000$&$0.824\pm0.030$\cr
		FGSM&$0.260\pm0.000$&$0.644\pm0.015$&$0.652\pm0.032$&$0.120\pm0.013$&$\bm{0.220\pm0.000}$&$0.516\pm0.037$\cr
		FGSM-ada&$0.240\pm0.000$&$0.592\pm0.016$&$0.664\pm0.034$&$\bm{0.116\pm0.015}$&$\bm{0.220\pm0.000}$&$\bm{0.504\pm0.029}$\cr
		AFGSM &$0.252\pm0.016$&$0.528\pm0.016$&$\bm{0.604\pm0.034}$&$0.156\pm0.008$&$0.224\pm0.008$&$0.728\pm0.030$\cr
		AFGSM-ada&$\bm{0.216\pm0.013}$&$0.460\pm0.013$&$0.656\pm0.028$&$0.136\pm0.020$&$0.224\pm0.023$&$0.648\pm0.030$\cr

		\bottomrule
	\end{tabular}
	}
\end{table*}

\begin{figure}
    \begin{minipage}{0.4\textwidth}
    	\makeatletter\def\@captype{table}\makeatother
    	\caption{Accuracy on Reddit}\label{tab:reddit}
    	\setlength{\tabcolsep}{2pt}
    	\scalebox{0.8}{
    	\begin{tabular}{lcc}
    		\toprule
    		\multirow{2}{*}{Method}&
    		\multicolumn{2}{c}{Reddit}\cr
    		\cmidrule(lr){2-3} 
    		&GCN&Deepwalk\cr
    		\midrule
   		Clean&$0.860\pm0.010$&$0.96\pm0.000$\cr
    	    Random&$0.860\pm0.042$&$0.944\pm0.008$\cr
    	    AFGSM&$\bm{0.572\pm0.027}$&$\bm{0.892\pm0.020}$\cr
    		\bottomrule
    	\end{tabular}}
	\end{minipage}
	\begin{minipage}{0.55\textwidth}
    \includegraphics[width=6.8cm]{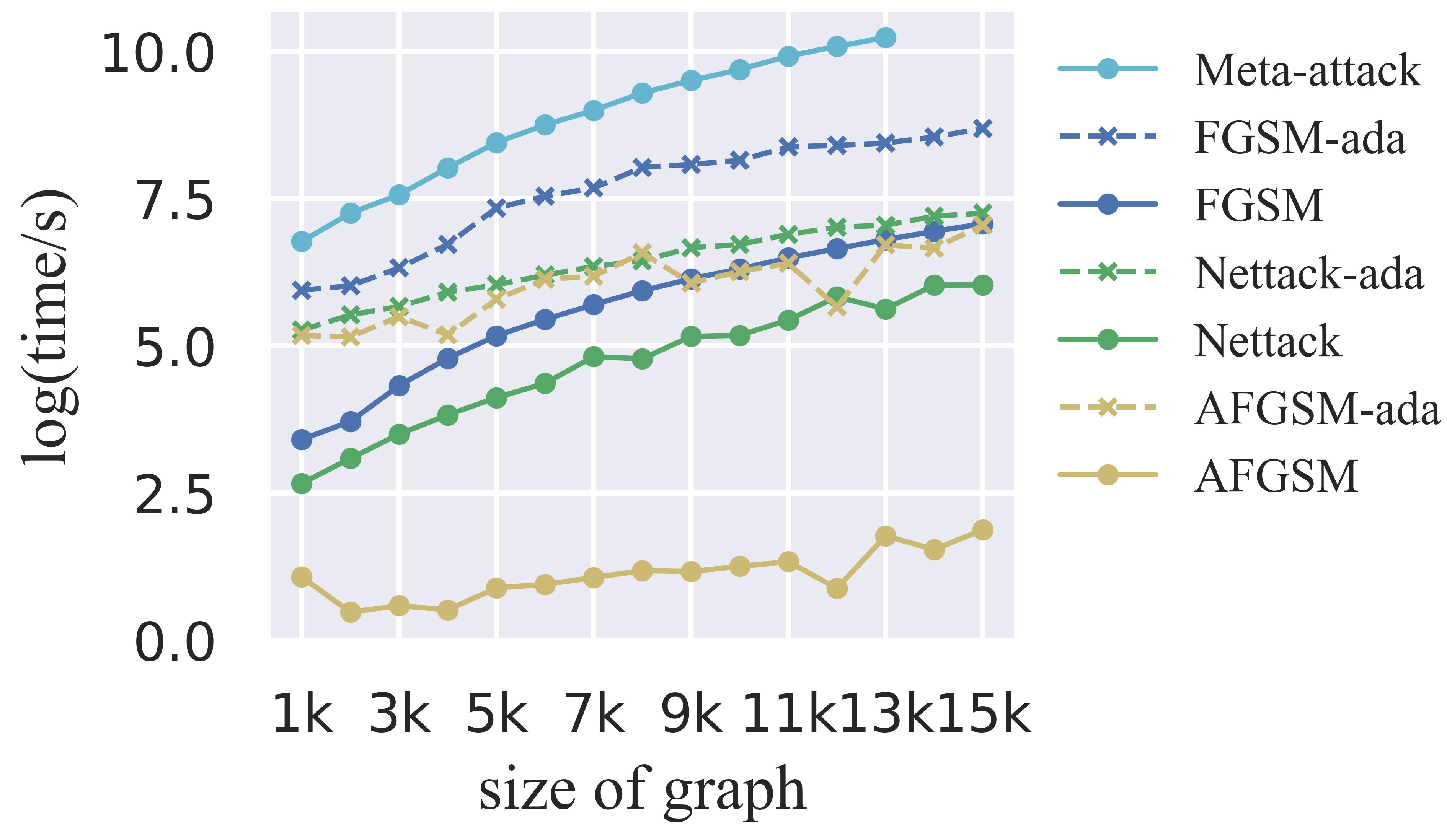}
    \caption{Run time comparison on the GCN model.}\label{fig:time_cost}
    \end{minipage}
\end{figure}

\subsection{Edge-only perturbation and Indirect perturbation}\label{sec:strict limits}
In this section, we explore two additional restricted perturbations: edges-only perturbation and indirect perturbation to verify the robustness of our attack algorithm in more practical and restricted settings. 
\begin{figure}[t] 
	\centering
	\includegraphics[width=0.9\columnwidth]{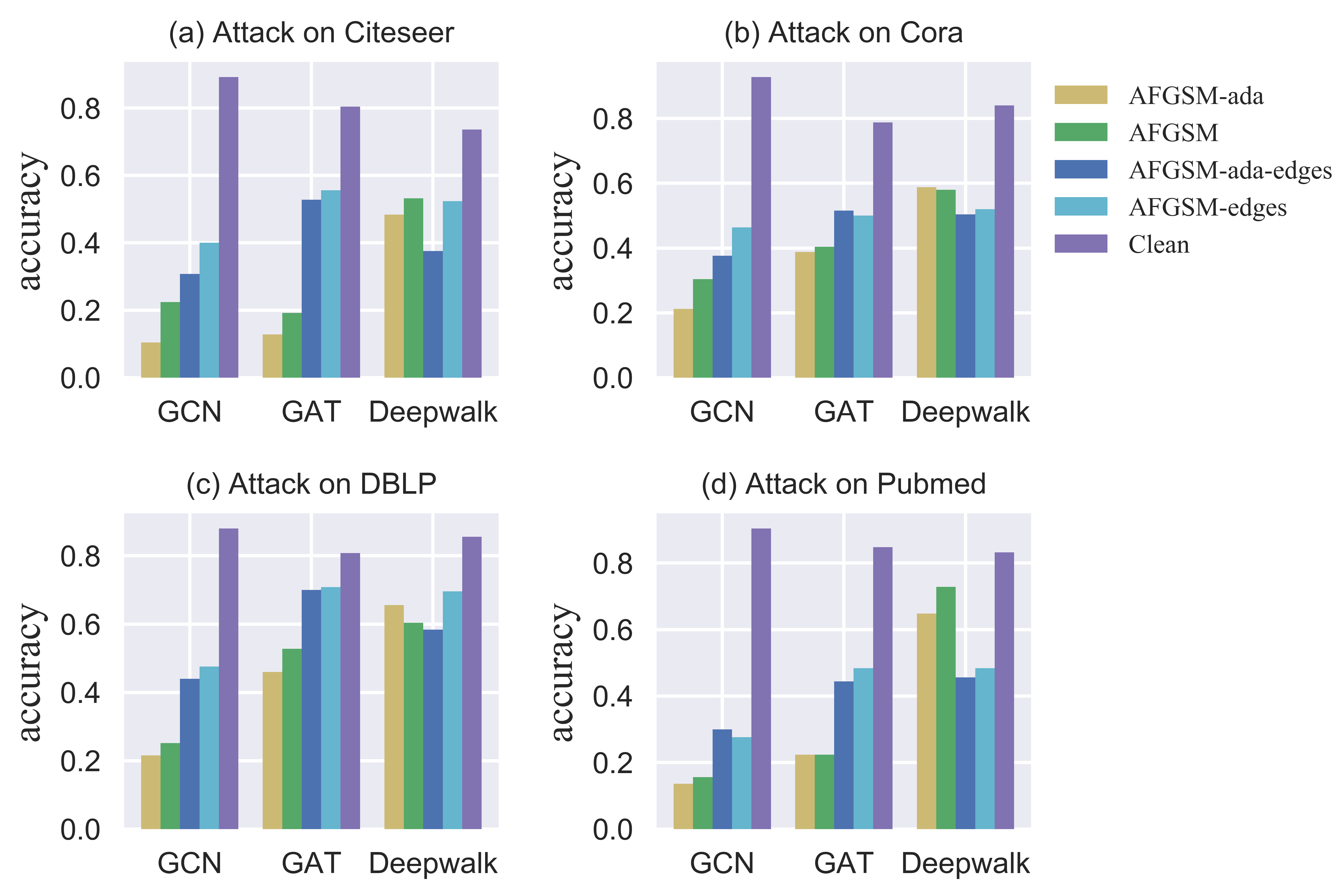}
	\caption{Accuracy of victim learning model against edge-only attacks.}\label{fig:edge-only}
\end{figure}

\begin{figure}[t] 
	
	\centering
    \includegraphics[width=0.9\columnwidth]{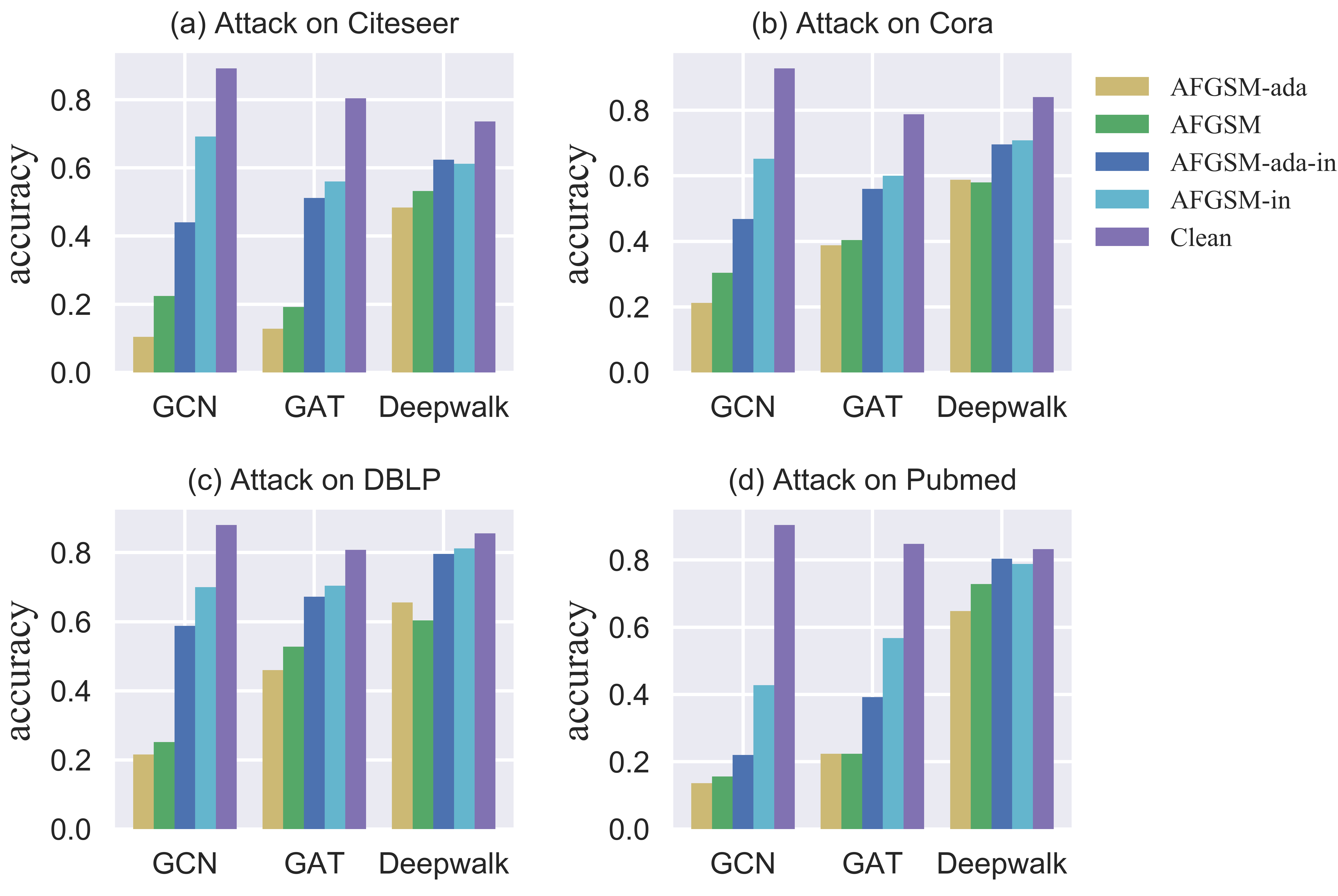}
	\caption{Accuracy of victim learning model against indirect attacks.}\label{fig:indirect}

\end{figure}

\paragraph{\bfseries{Edge-only perturbation.}} 
From the practical point of view, manipulation of features can be hard since attackers may not have the knowledge of how features in the graph are selected and preprocessed. Therefore, we study the performance AFGSM when attackers are only allowed to change edges of vicious nodes. To do so, we inject vicious nodes with random features sampled from the original graph instead of some well-designed features to get rid of the impact of features and only focus on the edges. We denote our attack in the restricted setting as AFGSM-edges since we only optimize edges during the attack process. 
The results are shown in Figure~\ref{fig:edge-only}. We observe that AFGSM in the restricted setting still succeeds. We further verify whether features and edges are equally important in the new attack scenario. By comparing the performance of AFGSM and AFGSM-edges as well as AFGSM-ada and AFGSM-ada-edges, we observe that perturbing only the edges limits the attack performance significantly. This is in contrast to the findings in the attack scenario of~\citep{zugner2018adversarial}, where the authors observe that manipulating features are not very important for successful attacks. Such contrast exists because, in the scenario of~\citep{zugner2018adversarial}, the attacker can only perturb features of the original nodes slightly (to remain unnoticeable) while in the new attack scenario, original features of vicious nodes can be rather arbitrary (but perturbations still follow the constraint in Eq.~(\ref{eq:constraint})). Hence, the new attack scenario grants the attacker more freedom in designing the perturbed features.

\paragraph{\bfseries{Indirect perturbation.}} In some cases, attackers may not be able to build connections with the target node directly. For example, some Facebook users can change their privacy settings and do not allow friend requests from unknown users who do share any mutual friends. Therefore, we evaluate the performance of our attack when attackers are not allowed to build direct connections to the target node. We denote the variant of our attack as AFGSM-in in the restricted setting. Results are shown in Figure~\ref{fig:indirect}. We observe that even in the restricted setting, the attack still succeeds in fooling the victim learning model. Unsurprisingly, we also observe that attacks in the current setting are much less successful than the attack in the case where vicious nodes can be directly connected to the target node. This observation also highlights the importance of building direct connections to the target node. 

\vspace{-4pt}
\section{Conclusion}
In this paper, we consider a practical attack scenario against GCN models, where attackers are only allowed to inject vicious nodes to the graph. We then propose a new attack algorithm named AFGSM to generate reliable adversarial perturbations efficiently. Through extensive experimental evaluations, we verify that GCN models are still vulnerable in the new attack setting and further show that our proposed AFGSM method outperforms the baselines significantly in terms of attack effectiveness and efficiency.

\small
\bibliographystyle{spbasic}    
\bibliography{sample-base}

\end{document}